\documentclass[11pt]{article}

\usepackage[margin=1in]{geometry}
\usepackage{microtype}

\usepackage{iftex}
\ifPDFTeX
  \usepackage[T1]{fontenc}
  \usepackage[utf8]{inputenc}
\else
  \usepackage{fontspec}
\fi

\usepackage{amsmath,amsfonts}
\usepackage{booktabs}

\usepackage{graphicx}
\usepackage{xcolor}

\usepackage{caption}   % you use \captionsetup
\usepackage{float}     % for [H] placement

\usepackage{enumitem}
\setlist[enumerate]{leftmargin=*,labelsep=0.5em}

\usepackage{siunitx}
\usepackage[authoryear,round]{natbib}

\usepackage{hyperref}
\hypersetup{
  colorlinks=true,
  linkcolor=black,
  citecolor=black,
  urlcolor=black,
  pdftitle={Negotiative Alignment: Embracing Disagreement for Fairer Urban Outcomes},
  pdfauthor={Rashid Mushkani, Hugo Berard, Shin Koseki}
}

\usepackage{authblk}

\newcommand{\keywords}[1]{\par\noindent\textbf{Keywords:} #1}

\providecommand{\Description}[2][]{}

\title{\vspace{-6pt}Negotiative Alignment: Embracing Disagreement to Achieve Fairer Outcomes -- Insights from Urban Studies\vspace{4pt}}

\author[1,2]{Rashid Mushkani}
\author[1]{Hugo Berard}
\author[1,2]{Shin Koseki}
\affil[1]{Universit\'e de Montr\'eal}
\affil[2]{Mila -- Quebec AI Institute}
\date{} % omit date

\begin{document}
\maketitle

\begin{abstract}
Urban assessments often compress diverse needs into single scores, which can obscure minority perspectives. We present a community-centered study in Montreal (n=35; wheelchair users, seniors, LGBTQIA2+ residents, and immigrants). Participants rated 20 streets (accessibility, inclusivity, aesthetics, practicality) and ranked 7 images on 12 interview-elicited criteria. Disagreement patterns were systematic in our sample: wheelchair users diverged most on accessibility and practicality; LGBTQIA2+ participants emphasized inclusion and liveliness; seniors prioritized security. Group discussion reduced information gaps but not value conflicts; ratings conveyed intensity, while rankings forced trade-offs. We then formalize \emph{negotiative alignment}, a transparent, budget-aware bargaining procedure, and pilot it with role-played stakeholder agents plus a neutral mediator. Relative to the best base design under the same public rubric, the negotiated package increased total utility (21.10 to 24.55), raised the worst-group utility (3.20 to 3.90), improved twentieth-percentile satisfaction (0.86 to 1.00; min–max normalized within the scenario), and reduced inequality (Gini 0.036 to 0.025). Treating disagreement as signal and reporting worst-group outcomes alongside totals may help planners and AI practitioners surface trade-offs and preserve minority priorities while maintaining efficiency.
\end{abstract}


\section{Introduction}\label{sec:introduction}
Urban environments are shared, yet people experience the same streets through distinct social, cultural, and political lenses. A wheelchair user in Montreal may perceive a continuous sidewalk as interrupted by a single missing curb cut, whereas an LGBTQIA2+ resident may read street murals or nearby businesses as cues of inclusion or exclusion. Contemporary urban analysis often emphasizes aggregate metrics (e.g., single overall ratings for aesthetics, accessibility, or safety) \citep{Mehta2014Evaluating,Varna2010}. These averages can be useful at city scale, but they can dilute the specific and sometimes urgent needs of marginalized groups \citep{Fainstein2010, CostanzaChock2020}.

Drawing on a multi-phase study of public space in Montreal \citep{Mushkani2025JUM_MontrealStreets}, we argue that apparent disagreements in urban evaluations are not merely noise. In our sample of 35 participants, divergences aligned with differences in lived experience, structural inequality, and cultural values. Combining semi-structured interviews, rating tasks, and a forced-ranking experiment, we observed persistent disagreement, particularly for accessibility, inclusivity, and practicality. As detailed in Section~\ref{sec:results}, factual clarifications during group discussion resolved some discrepancies, yet identity-grounded value differences remained.

Conventional survey and AI methods often prioritize consensus by aggregating heterogeneous preferences into a single score, which can obscure minority viewpoints \citep{Gabriel2020, Rawls1999}. Recent advances incorporate pluralistic modeling, game-theoretic reasoning, and fairness-aware optimization \citep{Zhi-Xuan2024, SorensenEtAl2024, Jain_2024,dwork2012fairness,Ferrara_2023}, but few include explicit, auditable mechanisms for negotiating conflicts among stakeholders during decision making. Addressing this gap calls for alignment strategies that iteratively adjust stakeholder weights and trade-offs rather than statically aggregating preferences.

We introduce \emph{negotiative alignment}, a dynamic, multi-agent approach that integrates disagreement into iterative decision making (Section~\ref{sec:negotiation}). Unlike singular consensus methods, negotiative alignment identifies persistent conflicts, distinguishes information gaps from value conflicts, and uses transparent rules to search for packages that raise the floor for the worst off while maintaining strong overall performance.

\subsection{Contributions and Research Questions}
We organize the paper around five questions that connect qualitative insight with quantitative analysis and an implementable negotiation workflow:
\begin{enumerate}
    \item \textbf{Who disagrees about what, and why?} We identify criteria (e.g., accessibility, inclusivity) that yield the strongest disagreements and analyze structural, cultural, and personal factors.
    \item \textbf{How do different evaluation methods capture disagreement?} We contrast ratings (which convey intensity) with rankings (which impose trade-offs) and examine how each reflects disagreement among intersectional populations \citep{Leroy2024,Xu2024,Rofin2023,SuzukiHorita2024,FritzMorettiStaudacher2023}.
    \item \textbf{When does group discussion narrow discrepancies, and when does it not?} We analyze focus-group interactions to separate information gaps from value conflicts and extract design principles for an algorithmic procedure.
    \item \textbf{How can negotiative alignment be incorporated into computational models?} We present a multi-agent framework that preserves minority viewpoints through explicit weights and floor-raising rules.
    \item \textbf{Can a mediated, role-played negotiation improve outcomes in a pilot?} Using stakeholder agents and a neutral mediator grounded in our interview data, we evaluate whether the negotiated package improves both worst-group and aggregate outcomes relative to the best base design.
\end{enumerate}
The remainder of the paper reviews related work, details the methodology, presents empirical results, describes the negotiation pilot and operator, and ends with limitations, ethics, and implications for practice.

\section{Related Literature}\label{sec:related}
Urban assessment frameworks frequently aggregate feedback into composite metrics for decision making \citep{Mehta2014Evaluating, Varna2010, naikstreetscore, qian2023aiagenturbanplanner, yurrita2021dynamicurbanplanningagentbased}. These techniques are efficient for policy and administration but can miss the nuanced needs of marginalized communities \citep{Fainstein2010, Sandercock1998,shreeyash}. Inaccessible sidewalks \citep{Imrie2012} and queer space erasure \citep{Doan2011} exemplify how aggregate measures can hide context-specific barriers. Planning toolkits such as the Gehl Institute’s public life tools \citep{Gehl2013} and the Project for Public Spaces Place Diagram \citep{SantosNouri2017}, together with indices such as the Public Space Index \citep{Mehta2014Evaluating} and the Pedestrian Environment Quality Index \citep{Zamanifard2019}, incorporate observation and engagement yet typically culminate in composite scores \citep{mushkani2025streetreviewparticipatoryaibased}. Critical urban studies warn that consensus-based measures can underrepresent minoritized needs. Our study complements these approaches by keeping disagreement explicit and turning persistent divergence into a resource for decision making.

\subsection{Alignment, Pluralism, and Negotiation}
Traditional alignment maps a model’s objective to a representative preference or an aggregate of preferences \citep{Gabriel2020, Rahwan2019machine}. Pluralistic methods move beyond a single consensus by modeling distributions of viewpoints \citep{Zhi-Xuan2024, SorensenEtAl2024,Nayak2024MIDSpace, Mushkani2025ICML_LIVS}. Multi-agent reinforcement learning and bargaining theory supply equilibrium and compromise tools \citep{zhang2021multi, yang2022learning, Nash1950, kalai1975, ChenHooker2023, rosen1965existence,Doostmohammadian2025}. Deliberative AI emphasizes structured dialogue and argumentation \citep{walton2009argumentation,rahwan2009,Nakamura2024,fish2023generativesocialchoice}. Fairness-conscious optimization sets floors or parity constraints across demographic groups \citep{moulin2004fair, dwork2012fairness,Nakamura2024}. Few of these lines specify a simple, auditable procedure that keeps minority priorities legible, tunes floors and weights over rounds, and exposes the trade-offs that drive convergence or stalemate.

\subsection{Positioning Negotiative Alignment}
Negotiative alignment augments pluralistic modeling with rule-bound bargaining steps that are visible to participants and evaluators. The approach is well suited to urban planning where infrastructure, demographics, and cultural meanings evolve and where stakeholder needs can compete. Past work simulates heterogeneous agents in planning \citep{qian2023aiagenturbanplanner, yurrita2021dynamicurbanplanningagentbased, zhou2024largelanguagemodelparticipatory, ni2024planninglivingjudgingmultiagent} but rarely formalizes negotiation for intractable conflicts. Dissent-aware learning highlights the importance of preserving minority voices \citep{Gordon_2022, Chen_2021} yet leaves open how to adjudicate incompatible interests once identified. Our operator fills this gap with a minimal set of transparent rules and metrics that can be inspected and audited.

\section{Methodology}\label{sec:method}
\subsection{Setting and Participants}
\begin{figure}[ht]
\centering \includegraphics[width=1\linewidth]{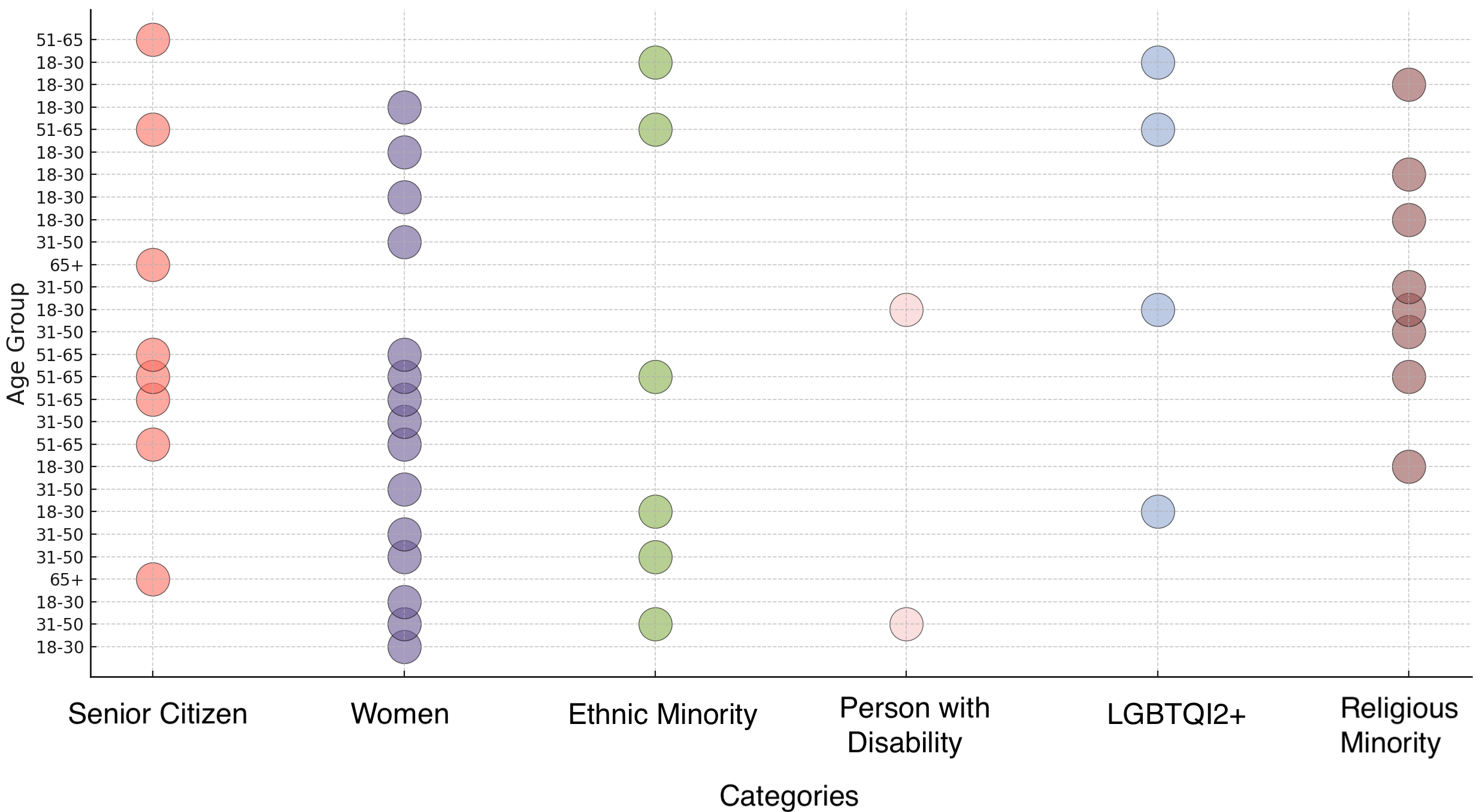}
\caption{Representation of diversity across age groups. The distribution of self-reported demographic characteristics varies by age, underscoring intersectional perspectives and cautioning against reducing any experience to a single identity \cite{mushkani2025promptcommonscollectiveprompting}.}
\captionsetup{skip=0pt}
\Description{Representation of diversity across age groups}
\label{fig:diversity-participants}
\end{figure}

The study took place in Montreal from summer 2023 to autumn 2024. The city’s linguistic diversity, varied architectural forms, and intersecting social communities supported recruitment across wheelchair users, seniors, LGBTQIA2+ residents, and immigrants \cite{Margier2013, Sylvestre2010}. Figure~\ref{fig:diversity-participants} summarizes age distribution and demographic intersections. Participants were recruited through outreach to more than one hundred community organizations and partnerships with over thirty institutions serving the focal groups \cite{IRCGM2018}. Across phases, the sample comprised thirty-five individuals: twenty-eight completed in-depth interviews; twelve took part in rating sessions conducted individually and in small groups; and seventeen completed a ranking experiment. Participants received modest compensation aligned with partner guidance. All procedures were approved by a university Research Ethics Board; informed consent was obtained; and data were anonymized at collection.

\subsection{Design and Instruments}
The research unfolded in three phases. Phase I used semi-structured interviews (40–90 minutes) to elicit definitions of quality in public spaces, including usage patterns, meaningful places, difficult or exclusionary experiences, and desired qualities \cite{Creswell2022}. The interviews surfaced a vocabulary of criteria that later structured the ranking experiment \citep{MushkaniKoseki2025Habitat_StreetReview}. Phase II used three focus groups of three to five participants. Each group independently rated twenty street images on a four-point scale—where 1 indicated poor and 4 indicated excellent—for inclusivity, accessibility, aesthetics, and practicality. Street views were selected to cover a socio-spatial matrix varying land use, density, greenery, historical character, and design elements (see Appendix Figures~\ref{fig:street_diversity_classification}, \ref{fig:street_views}, and \ref{fig:spatial_distribution}). Individual ratings were followed by a structured discussion in which participants compared justifications, identified information gaps (e.g., unnoticed ramps or tactile paving), and articulated value differences (e.g., comfort with crowds or symbolic inclusion) \cite{Margier2013}. Phase III asked seventeen participants to rank seven images from best to worst for twelve criteria elicited from interviews: accessibility, invitingness, comfort, regenerativeness, beauty, practicality, maintenance, inclusivity, dynamism, representativeness, oppression, and security. We reduced the image set from twenty to seven to enable deeper discussion while maintaining variation across the socio-spatial matrix.

Image presentation was randomized to mitigate order effects. For ratings, anchors were defined and printed on a reference card for each criterion. For rankings, ties were disallowed to make trade-offs explicit. Participants could request magnified views and clarifying descriptions of visible features, yet no additional context beyond the images was provided to avoid privileging place familiarity. Sessions were audio recorded and transcribed with consent.

\subsection{Analysis and Validity}
Qualitative analysis used thematic coding to explain why people disagreed, with particular attention to mobility constraints, symbolic inclusion, and safety \cite{Creswell2022}. Codes that emerged from interviews informed the criteria used in subsequent tasks, and memos documented analytic decisions. A subset of transcripts was double coded by two researchers who discussed disagreements until consensus and updated the codebook. This process emphasized interpretive reliability without reducing the analysis to mechanical interrater statistics.

Quantitative analysis treated ratings as ordinal and used rank-based correlations by default. We computed Kendall’s Tau to quantify ordinal agreement across groups for each criterion \cite{Abdi2006}. For diagnostic comparison with prior work, we also report Pearson correlations for compiled ratings aggregated at the image level \citep{Puth2015Effective}. Where appropriate, we analyzed differences across groups using nonparametric tests such as Kruskal–Wallis and pairwise Wilcoxon comparisons with Benjamini–Hochberg correction, and we report effect sizes based on Cliff’s delta rather than relying on significance alone. For rankings we assessed within-group concordance using Kendall’s W. Confidence intervals were estimated by nonparametric bootstrap. Our goal was to characterize the structure and persistence of disagreement rather than to claim population-level effects.

\subsection{Ethics, Community Partnership, and Data Governance}
The study adopted a co-production philosophy with community partners \citep{Mushkani2025AIES_CoProducingAI}. Organizations advised on recruitment practices and accessibility, and we provided feedback sessions and materials in return \citep{IRCGM2018,Mushkani2025ICML_RightToAI}. Consent forms used plain language. Participants could pause or withdraw without penalty. Data were stored on encrypted drives with restricted access. Quotations used in the paper were anonymized and screened for identifiability. Because place familiarity can reveal identity, we avoided naming or showing exact locations during group discussion and masked signage in public images when feasible. Consistent with this policy, examples in Results use anonymous site codes.

\subsection{Metrics and Notation for Decision Support}
Negotiative alignment requires clear, auditable metrics. Let $G$ denote the set of stakeholder groups and $K$ the set of criteria. A design package $x$ has attribute scores $s_k(x)$ on a five-point scale for each $k \in K$. Group $g \in G$ has normalized, nonnegative weights $w_{gk}$ that sum to one across $k$. Group utility is $U_g(x)=\sum_{k \in K} w_{gk} s_k(x)$. We report four aggregate metrics. Worst-group utility is $W(x)=\min_{g \in G} U_g(x)$. The utilitarian sum is $S(x)=\sum_{g \in G} U_g(x)$. The twentieth-percentile satisfaction is the empirical 0.20 quantile of $\{U_g(x)\,:\,g \in G\}$, min–max normalized to $[0,1]$ within the scenario’s feasible set for comparability. Inequality is summarized by the Gini coefficient over group utilities,
\[
G(x)=\frac{\sum_{g}\sum_{g'}|U_g(x)-U_{g'}(x)|}{2|G|\sum_{g}U_g(x)}.
\]
These definitions enable transparent comparison of baselines and negotiated outcomes.

\section{Results}\label{sec:results}
\subsection{Overview of Street-Level Findings}
\begin{figure}[ht]
    \centering
    \includegraphics[width=1\linewidth]{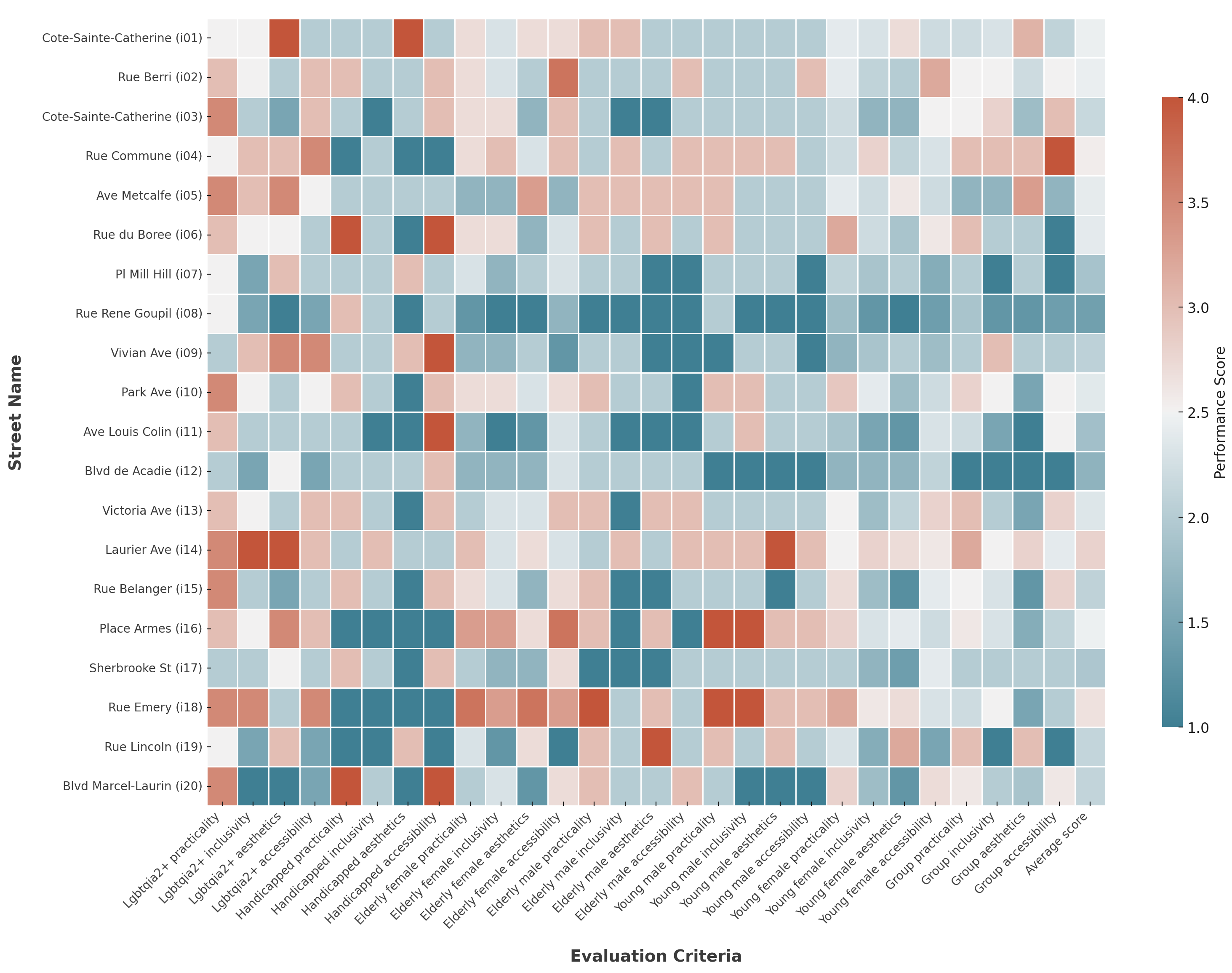}
    \caption{Matrix of compiled ratings for twenty Montreal streets across inclusivity, accessibility, aesthetics, and practicality, disaggregated by core demographic groups. Distinct and persistent disagreements are visible.}
    \captionsetup{skip=0pt}
    \Description{Matrix of compiled ratings for 20 Montreal streets}
    \label{fig:all-evals}
    \vspace{-15pt}
\end{figure}

Figure~\ref{fig:all-evals} summarizes aggregated ratings for twenty streets across inclusivity, accessibility, aesthetics, and practicality, disaggregated by core demographic groups. Persistent and patterned disagreements emerged. The most pronounced divergences occurred for accessibility and practicality among wheelchair users and for inclusivity and liveliness cues among LGBTQIA2+ participants. These patterns are consistent with the view that disagreement encodes structurally patterned needs rather than idiosyncratic noise.

Three anonymized cases illustrate how divergences arise and why averages can mislead. Site \textbf{i08} received a low aggregate score of 1.56, largely because multiple groups rated inclusivity as 1.0. Site \textbf{i16} revealed a sharp split on accessibility: wheelchair users assigned accessibility a score of 1.0 due to curb obstructions that other groups did not notice or did not consider decisive. Site \textbf{i14} drew favorable assessments for aesthetics and inclusivity yet was criticized for accessibility by participants with mobility impairments. On site \textbf{i06}, a wheelchair user rated accessibility as 4.0 in recognition of ramps and surface quality, whereas some LGBTQIA2+ participants rated aesthetics as 1.0 and expressed concerns about symbolic inclusivity and safety—a reminder that function and perceived welcome can pull in opposite directions.

\subsection{Who Disagrees, and Why}
Wheelchair users diverged most on accessibility and practicality. Streets that presented well in photographs sometimes concealed traversability issues that were immediately salient to those with mobility experience. LGBTQIA2+ participants weighed symbolic inclusion, signage, and the character of nearby businesses more heavily than other groups and favored lively atmospheres when they communicated safety. Seniors were more attentive to noise, crowding, and lighting, factors they associated with perceived safety. These differences were not reducible to single identities. For example, younger wheelchair users were more accepting of lively streets than seniors in the same group but remained uncompromising on curb cuts and tactile paving.

Function and appeal traded off in several locations. Correlations showed near-zero or negative relationships between practicality and aesthetics at the street level, with cobblestones a recurring example. While visually appealing to many, cobblestones can be difficult to traverse in a wheelchair and uncomfortable for people who rely on canes or walkers. Ranking tasks magnified these tensions because they required participants to choose which criterion mattered more in a given context rather than allowing partial credit across all.

Familiarity mattered. Perceptions of inclusivity and aesthetics were shaped by lived experience and subcultural knowledge. Public art that some participants viewed as beautification was interpreted by others as a cue of gentrification or an unwelcome signal. Wheelchair users familiar with accessibility infrastructure responded positively to tactile paving even when it was not immediately obvious in images, whereas unfamiliar observers undervalued such features.

\begin{figure}[htbp]
    \centering
    \includegraphics[width=1\linewidth]{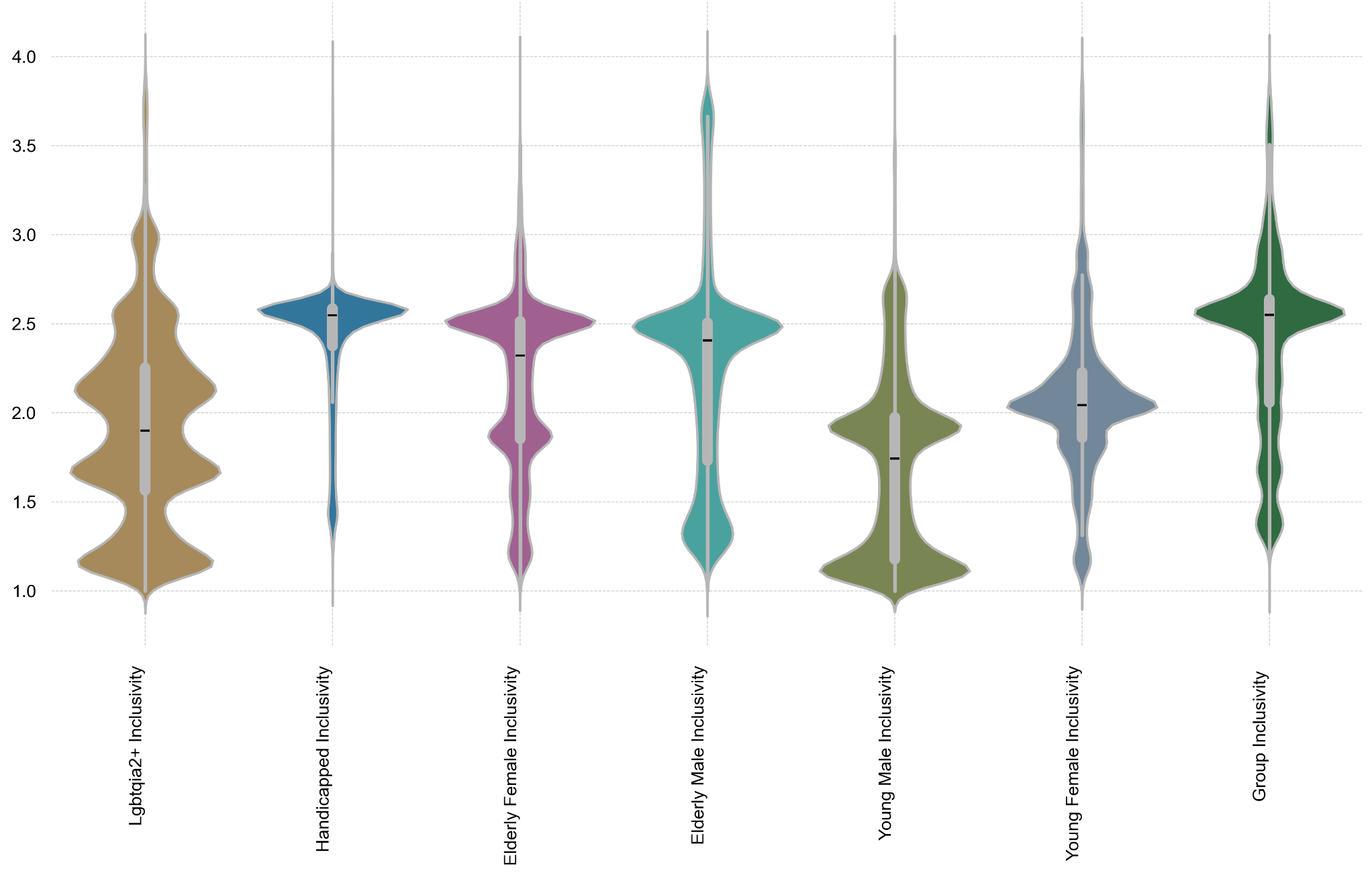}
    \caption{Violin plot of inclusivity ratings by demographic group. The distributions highlight differences in perceived inclusivity and the influence of features such as seating, lighting, and sidewalk conditions.}
    \Description{Violin Plot of Inclusivity Ratings}
    \label{fig:violin_plot}
    \vspace{-15pt}
\end{figure}

\subsection{Negotiation Effects in Group Settings}
Group discussion reduced discrepancies when they stemmed from information gaps. Participants sometimes revised ratings after a peer pointed out a ramp, a dropped curb, or a seating option. This calibration did not erase value-based disagreements that arose from identity-grounded needs and risk perceptions. The pattern supports a two-stage approach: first address information asymmetries; then recognize that some conflicts are principled and must be negotiated rather than averaged away.

\subsection{Ratings and Rankings Provide Complementary Signals}
The four-point rating scale supported partial convergence on less-contested dimensions and conveyed intensity of feeling. Rankings imposed explicit trade-offs and exposed where agreement was unattainable without redesign. This complementarity echoes insights from social choice research: ratings and rankings carry distinct information and should be used jointly when the goal is to preserve minority priorities while seeking workable compromises \cite{Leroy2024,Xu2024,Rofin2023,SuzukiHorita2024,FritzMorettiStaudacher2023}.

\subsection{Correlation Structure and Persistent Disagreement}
\begin{figure}[ht]
\centering
    \includegraphics[width=1\linewidth]{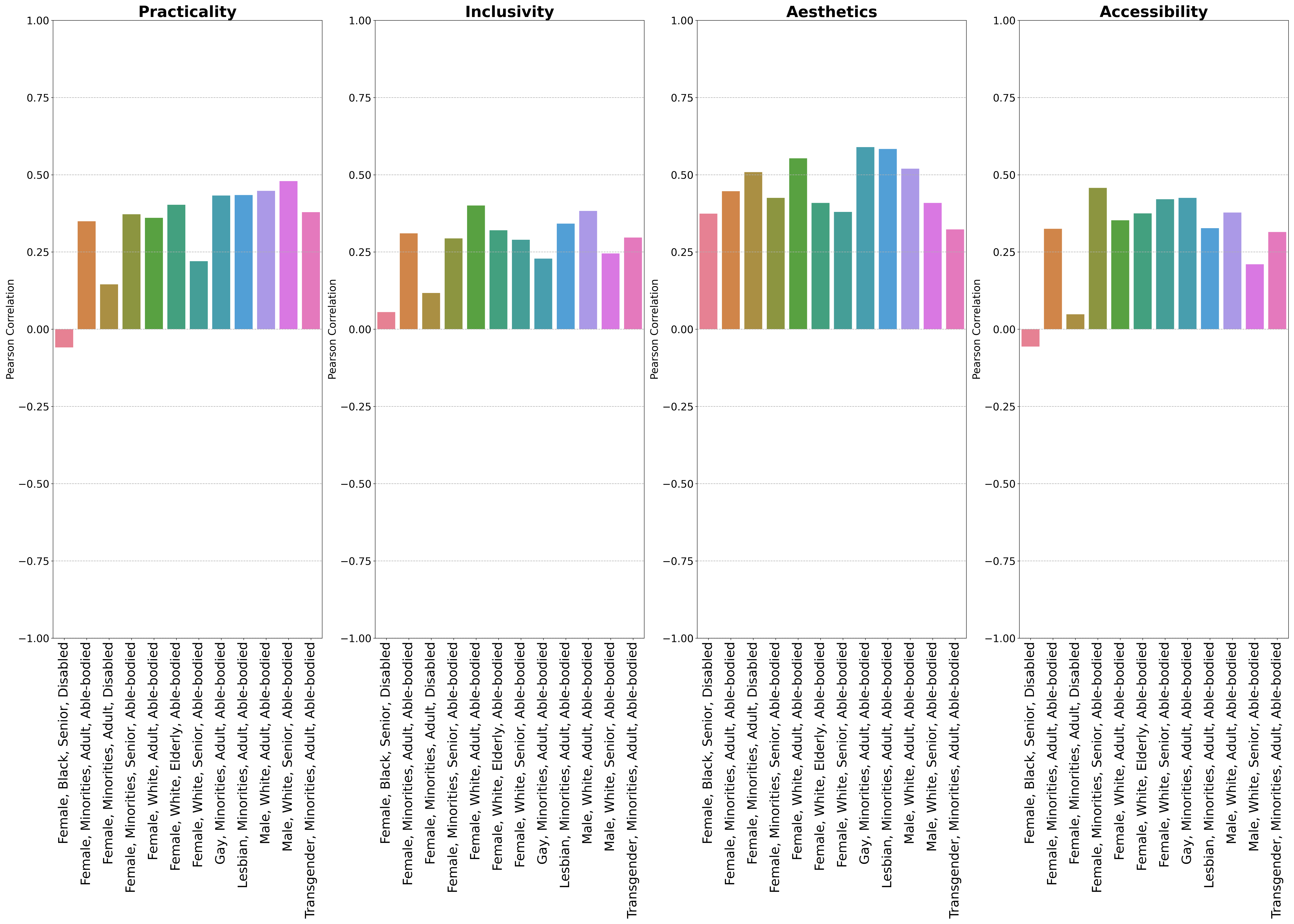}
    \caption{Ratings. Pearson correlations between demographic groups for practicality, inclusivity, aesthetics, and accessibility. Disagreement is most pronounced for accessibility and practicality among intersectional identities.}
    \Description{Rating using Pearson correlation}
    \label{fig:visualization}
    \vspace{-15pt}
\end{figure}

\begin{figure}[ht]
\centering
\includegraphics[width=1\linewidth]{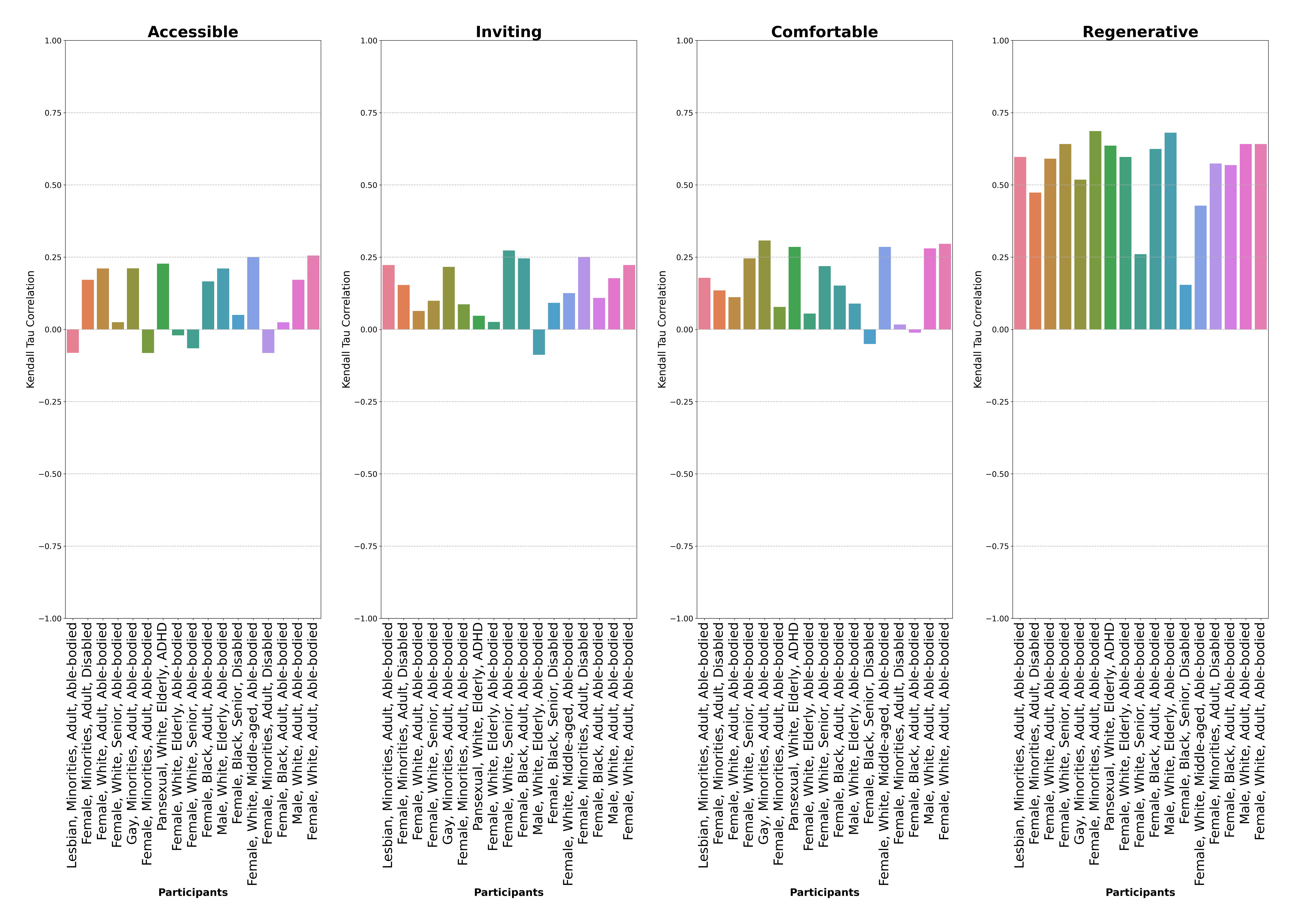}
\caption{Rankings. Kendall’s Tau for accessibility, invitingness, comfort, and regenerativeness. Regenerativeness shows higher consensus, while accessibility remains contentious.}
\Description{Ranking using Kendall Tau3}
\label{fig:kendall_tau_group_1}
\vspace{-15pt}
\end{figure}

\begin{figure}[ht]
\centering
\includegraphics[width=1\linewidth]{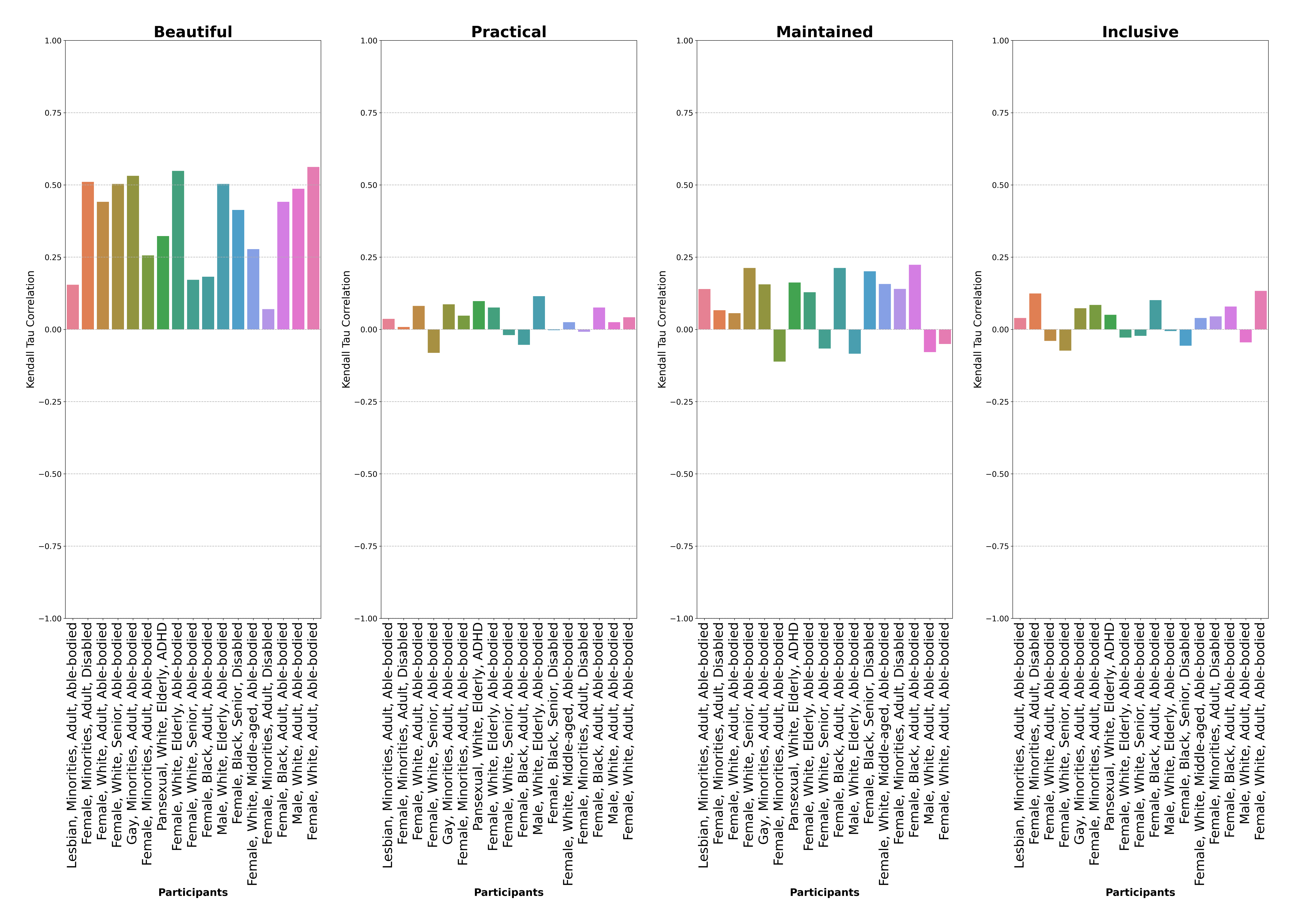}
\caption{Rankings. Kendall’s Tau for beauty, practicality, maintenance, and inclusivity. Disagreement is strongest for practicality and inclusivity.}
\Description{Ranking using Kendall Tau2}
\label{fig:kendall_tau_group_2}
\vspace{-15pt}
\end{figure}

\begin{figure}[ht]
\centering
\includegraphics[width=1\linewidth]{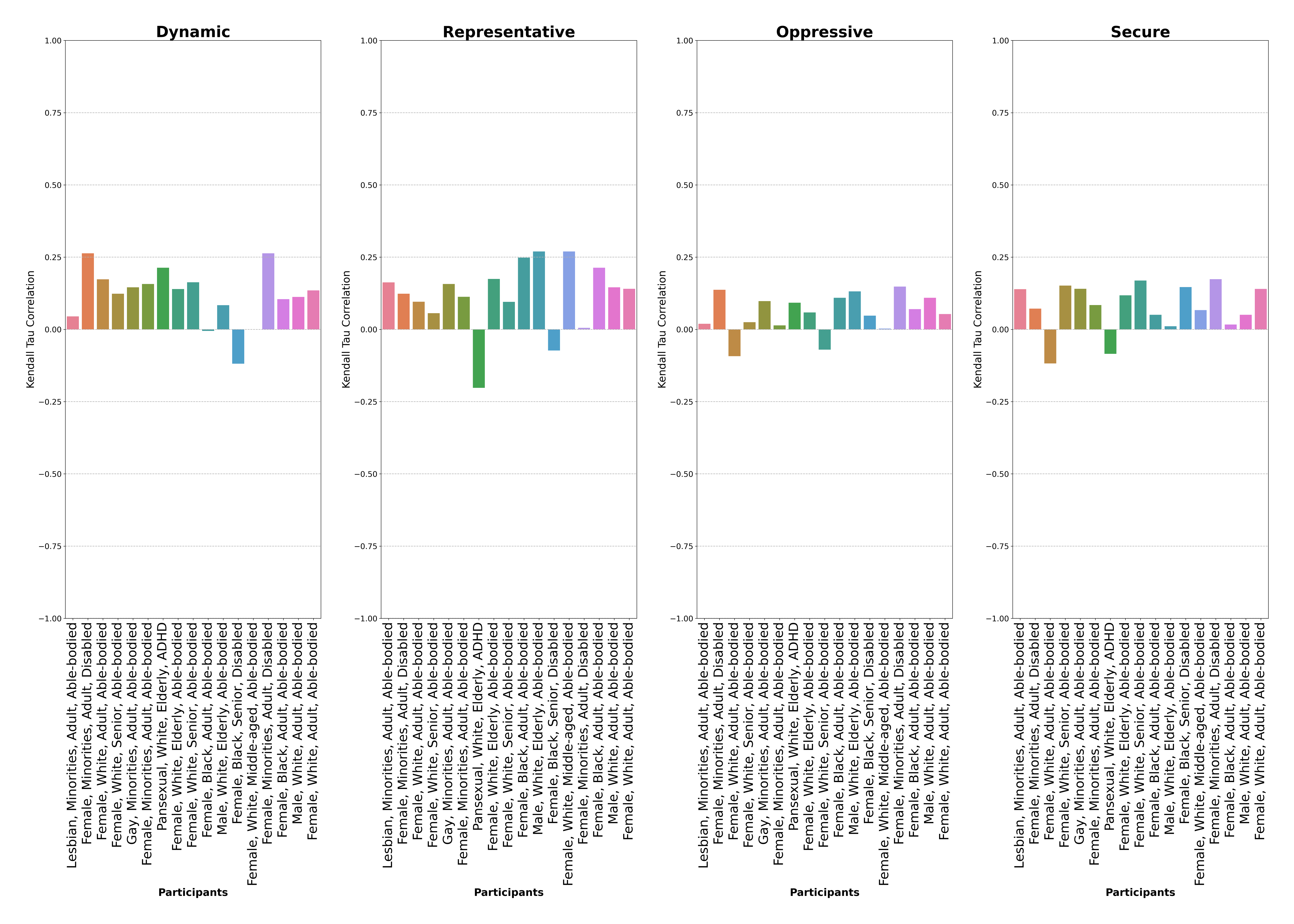}
\caption{Rankings. Kendall’s Tau for dynamism, representativeness, oppression, and security. These dimensions polarized participants and reflect divergent perceptions of safety, vibrancy, and cultural representation.}
\Description{Ranking using Kendall Tau}
\label{fig:kendall_tau_group_3}
\vspace{-15pt}
\end{figure}

Figures~\ref{fig:visualization} through~\ref{fig:kendall_tau_group_3} report correlation metrics across criteria. Pearson correlations between groups were higher for practicality and accessibility than for inclusivity and aesthetics, consistent with the idea that functional constraints sometimes admit technical fixes while symbolic inclusion and safety perceptions reflect social meaning. Kendall’s Tau plots for rankings show relatively higher consensus for regenerativeness and more disagreement for practicality and inclusivity. Security, dynamism, and representativeness polarized participants, with divergence tied to age and identity.

\begin{figure}[ht]
\centering
\includegraphics[width=1\linewidth]{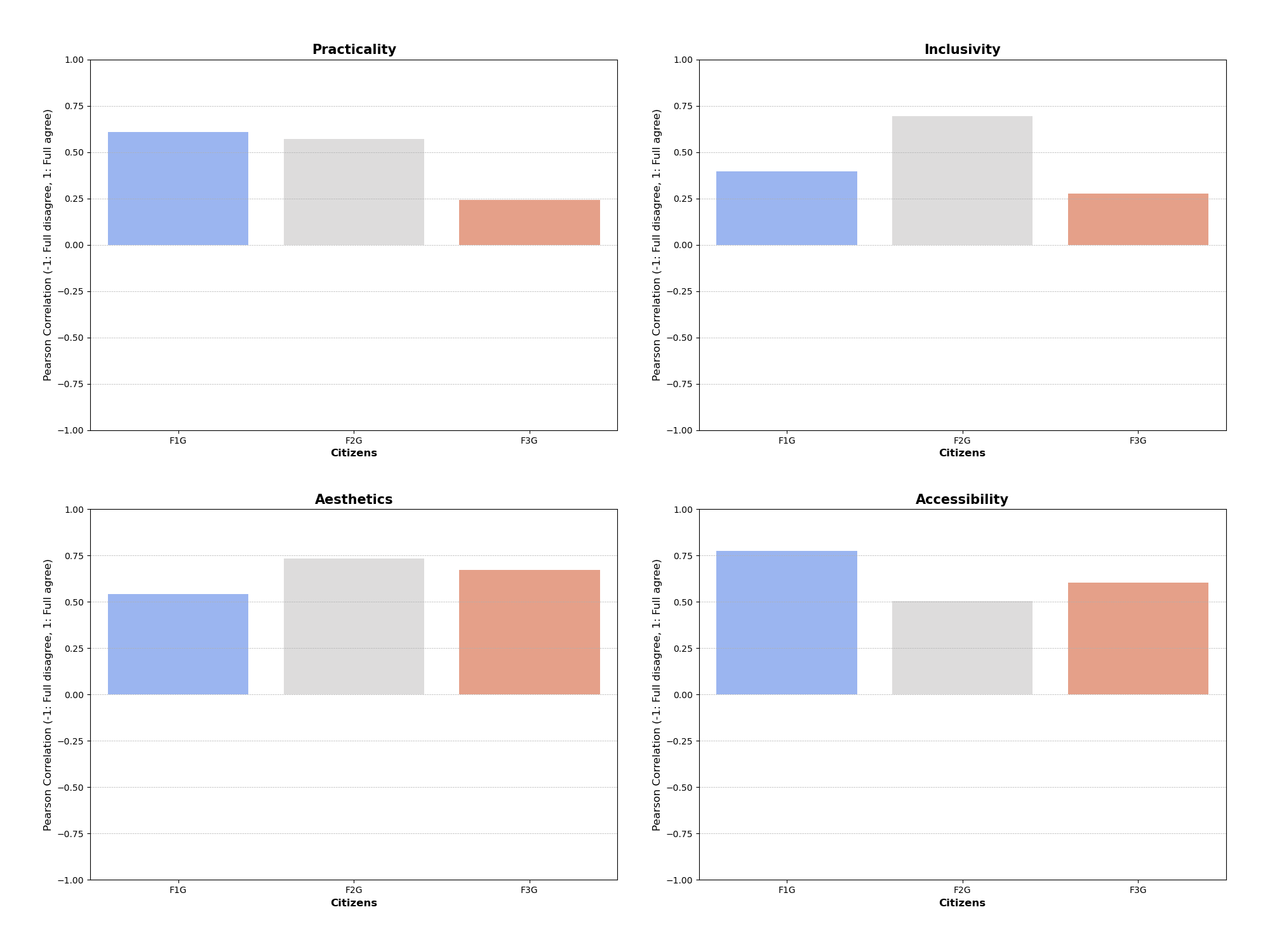}
\caption{Group ratings. Pearson correlations for selected subgroups (labels anonymized for review). The figure suggests more balanced evaluations within subgroups yet persistent divergence across groups.}
\Description{Group Rating}
\label{fig:detail_pearson_corr_rating_group}
\vspace{-5pt}
\end{figure}

\section{From Theory to Practice: An LLM-Agent Negotiation Experiment}\label{sec:negotiation}
We ask whether large language models can act as role-played stakeholder agents, together with a neutral mediator, to negotiate urban design decisions that improve the worst-group outcome while preserving minority priorities and maintaining strong aggregate performance. The experiment ties to the Montreal study by deriving agent roles and weights from interview themes and by constraining choices to street-design packages that reflect observed trade-offs.

\subsection{Personas, Utilities, and Scenarios}\label{subsec:personas}
Six role-played stakeholder agents represent salient identities and priorities: a wheelchair user (W), an LGBTQIA2+ resident (Q), senior women and men (SF, SM), and younger women and men (YF, YM). Each agent \(g\) evaluates a design package \(x\) via an auditable cardinal utility
\[
U_g(x)=\sum_{k \in \{\text{Acs, Inc, Aes, Pra, Sec}\}} w_{gk}\, s_k(x),
\]
where \(s_k(x)\) are public attribute scores and \(w_{gk}\) are group-specific weights derived from interview coding (Appendix~\ref{app:personas}). Scenarios comprise three base designs and a budgeted menu of improvements with specified costs and score deltas for accessibility, inclusivity, aesthetics, practicality, and security (Appendix~\ref{app:scenario-cards}). Base scores for the pilot appear in Table~\ref{tab:pilot-roles-scores}.

In the pilot, all stakeholder agents and the mediator used a contemporary large language model with tools and browsing disabled. The model operated solely over the public rubric, scenario cards, and prompts in Appendix~\ref{app:prompts}; decoding parameters are also reported there.

\begin{table}[H]
\centering
\small
\begin{tabular}{lccccc}
\toprule
\textbf{Base Design} & \textbf{Acs} & \textbf{Inc} & \textbf{Aes} & \textbf{Pra} & \textbf{Sec} \\
\midrule
A (lively, cobblestone, few ramps) & 2 & 3 & 4 & 2 & 3 \\
B (functional upgrades)            & 4 & 3 & 2 & 4 & 4 \\
C (lively with some ramps)         & 3 & 4 & 4 & 3 & 2 \\
\bottomrule
\end{tabular}
\caption{Pilot scenario: base design attribute scores.}
\label{tab:pilot-roles-scores}
\end{table}

\subsection{Negotiation Operator and Protocol}\label{subsec:protocol}
Negotiative alignment uses a mediator bound to a public rubric and budget. The operator begins from a feasible set of packages and proceeds in four steps:
\begin{enumerate}
    \item Each agent submits a base design and up to two improvements, with one-sentence justifications tied to role priorities.
    \item The mediator enumerates feasible packages under the budget, computes group utilities from public cards, and proposes the package that lexicographically maximizes the minimum utility and then the sum.
    \item The mediator publishes the package scores with a short explanation of how the floor was raised and why the option is efficient given the budget.
    \item Agents ratify or object. Any objection must propose a feasible alternative and show how it improves the minimum without reducing the sum by more than a small tolerance; otherwise the mediator’s proposal stands.
\end{enumerate}
The process is short by design and creates an audit trail that community partners and evaluators can review. For reproducibility, we provide the mediator rule, prompts, and hyperparameters in Appendix~\ref{app:prompts}. The global system prompt instructs agents to remain civil, follow role cards, and provide concise public reasons without revealing hidden chain of thought.

\subsection{Baselines, Metrics, and Pilot}\label{subsec:baselines-metrics}
We compare the negotiated outcome to the best base design computed from the same public rubric. Metrics are the four defined in Section~\ref{sec:method}: the minimum across groups, the sum across groups, the twentieth-percentile satisfaction (min–max normalized within the scenario), and the Gini coefficient of group utilities. The pilot used a budget of four units and an improvement menu consisting of ramps with tactile paving, benches with shade, lighting upgrades, and inclusive signage with art. The mediator selected the package that started from base design B and funded ramps with tactile paving together with inclusive signage and art. The resulting attributes were Acs=5, Inc=4, Aes=3, Pra=4, and Sec=4.

\paragraph{Utilities and Fairness Metrics.}
\begin{table}[H]
\centering
\small
\begin{tabular}{lcccc}
\toprule
\textbf{Group} & \textbf{A} & \textbf{B} & \textbf{C} & \textbf{$D^\star$} \\
\midrule
W  & 2.35 & 3.80 & 3.00 & 4.45 \\
Q  & 3.00 & 3.20 & 3.40 & 3.90 \\
SF & 2.65 & 3.70 & 2.85 & 4.15 \\
SM & 2.60 & 3.70 & 2.90 & 4.15 \\
YF & 2.95 & 3.25 & 3.30 & 3.90 \\
YM & 2.70 & 3.45 & 3.20 & 4.00 \\
\midrule
\textbf{Sum} & 16.25 & \textbf{21.10} & 18.65 & \textbf{24.55} \\
\textbf{Min across groups} & 2.35 & \textbf{3.20} & 2.85 & \textbf{3.90} \\
\bottomrule
\end{tabular}
\caption{Pilot utilities under the public rubric. The negotiated package $D^\star$ (B plus ramps with tactile paving and inclusive signage with art) improves both the sum and the minimum relative to the best base.}
\label{tab:pilot-utilities}
\end{table}

\begin{table}[H]
\centering
\small
\begin{tabular}{lcc}
\toprule
\textbf{Metric} & \textbf{Best Base (B)} & \textbf{Negotiated $D^\star$} \\
\midrule
Worst-group utility & 3.20 & \textbf{3.90} \\
Sum of utilities & 21.10 & \textbf{24.55} \\
P20 satisfaction (normalized) & 0.86 & \textbf{1.00} \\
Gini (across group utilities)  & 0.036 & \textbf{0.025} \\
\bottomrule
\end{tabular}
\caption{Fairness and process metrics. Under the public rubric, $D^\star$ raises the floor, increases the total, improves the twentieth percentile (min–max normalized within the scenario), and reduces inequality.}
\label{tab:pilot-metrics}
\end{table}

Substantively, the wheelchair user insisted on ramps with tactile paving; the LGBTQIA2+ resident emphasized inclusive signage and art; seniors supported ramps; younger participants supported inclusive signage and art and valued additional seating. The mediator proposed the pair of improvements that jointly raised the floor while maintaining efficiency within the budget. Appendix~\ref{app:transcript} includes an eight-turn excerpt illustrating brevity and clarity.

\subsection{Properties and Design Rationale}
Negotiative alignment is intentionally simple. The lexicographic objective gives priority to raising the floor and then maximizes the sum, which avoids selections that help the average while leaving one group behind. Under the stated assumptions (nonnegative weights and nonnegative score deltas capped at five), two monotonicity properties hold: (i) increasing the budget weakly increases the minimum utility; and (ii) increasing any criterion score for a package cannot reduce any group’s utility. These properties do not guarantee global optimality, yet they support explainability and auditability—key in participatory contexts.

\subsection{Ethics and Governance of LLM Mediation}
LLM-mediated bargaining is meant to augment, not replace, human deliberation \citep{mushkani2025wedesigngenerativeaifacilitatedcommunity, mushkani2025collectiverecoursegenerativeurban}. All rules, rubrics, and transcripts should be published and subject to community veto when synthesized packages conflict with lived realities \citep{mushkani2025urbanaigovernanceembed}. The mediator must enforce civility, forbid identity-denigrating claims, and constrain reasoning to public information. Our prompts, guardrails, and checks are documented in Appendix~\ref{app:prompts}. Although the pilot surfaced no violations under these guardrails, bias and failure modes remain possible and warrant continuous oversight by human facilitators.

\section{Limitations, Implications, and Future Work}
This study advances a case for treating disagreement as a first-class signal and offers an auditable operator that turns persistent divergence into a tractable bargaining problem. Several limitations shape interpretation. The sample of thirty-five participants in Montreal is context specific and sized for depth rather than population inference. Quantitative analyses therefore emphasize effect sizes and correlation structure over null-hypothesis testing. Visual tasks use images rather than in-situ evaluation, which limits sensory context and can understate place familiarity effects. The LLM pilot is a proof of concept that demonstrates feasibility and transparency rather than a definitive evaluation of negotiation protocols at scale. Persona weights are derived from interviews and can drift if not periodically refreshed with new data. Rubric misspecification and proxy gaps can bias outcomes. Mediation rules assume balanced influence across roles, whereas real-world processes are shaped by power and history.

These limits suggest several directions for future work. Scaling to additional scenarios and neighborhoods will test whether the observed disagreement patterns persist across contexts. Re-weighting personas from fresh interviews and introducing cross-checks from independent facilitators can reduce drift. Multi-mediator juries and alternative bargaining rules (e.g., Kalai–Smorodinsky or constrained egalitarian solutions) could be compared within the same harness. Human-subject evaluations can assess perceived fairness and legitimacy of mediated outcomes. Online extensions that incorporate streaming feedback from residents could adapt floors and weights in real time, which will require careful attention to gaming and governance. Finally, embedding the operator in municipal workflows will require interfaces that expose trade-offs, allow for veto rights by community partners, and log rationales for future audits.

\section{Conclusion}
Disagreement is not noise; it encodes structurally patterned needs central to equitable design. The Montreal study shows how accessibility, inclusivity, aesthetics, practicality, security, and symbolism can diverge across identities, and how group discussion clarifies information gaps without erasing value conflicts. By operationalizing negotiation with a rule-bound mediator under a public rubric, we show a practical path that improves the worst-group outcome and the overall sum while reducing inequality relative to the best base design. The workflow is simple, auditable, and compatible with participatory practice. It invites planners and communities to make disagreements visible, bargain over concrete packages, and measure what changes.

\bibliographystyle{plainnat}
\bibliography{0-base}

\appendix
\section{Supplementary Material}

\subsection{Heatmap of Correlations by Demographic Group}
\begin{figure*}[htbp]
    \centering
    \includegraphics[width=\textwidth]{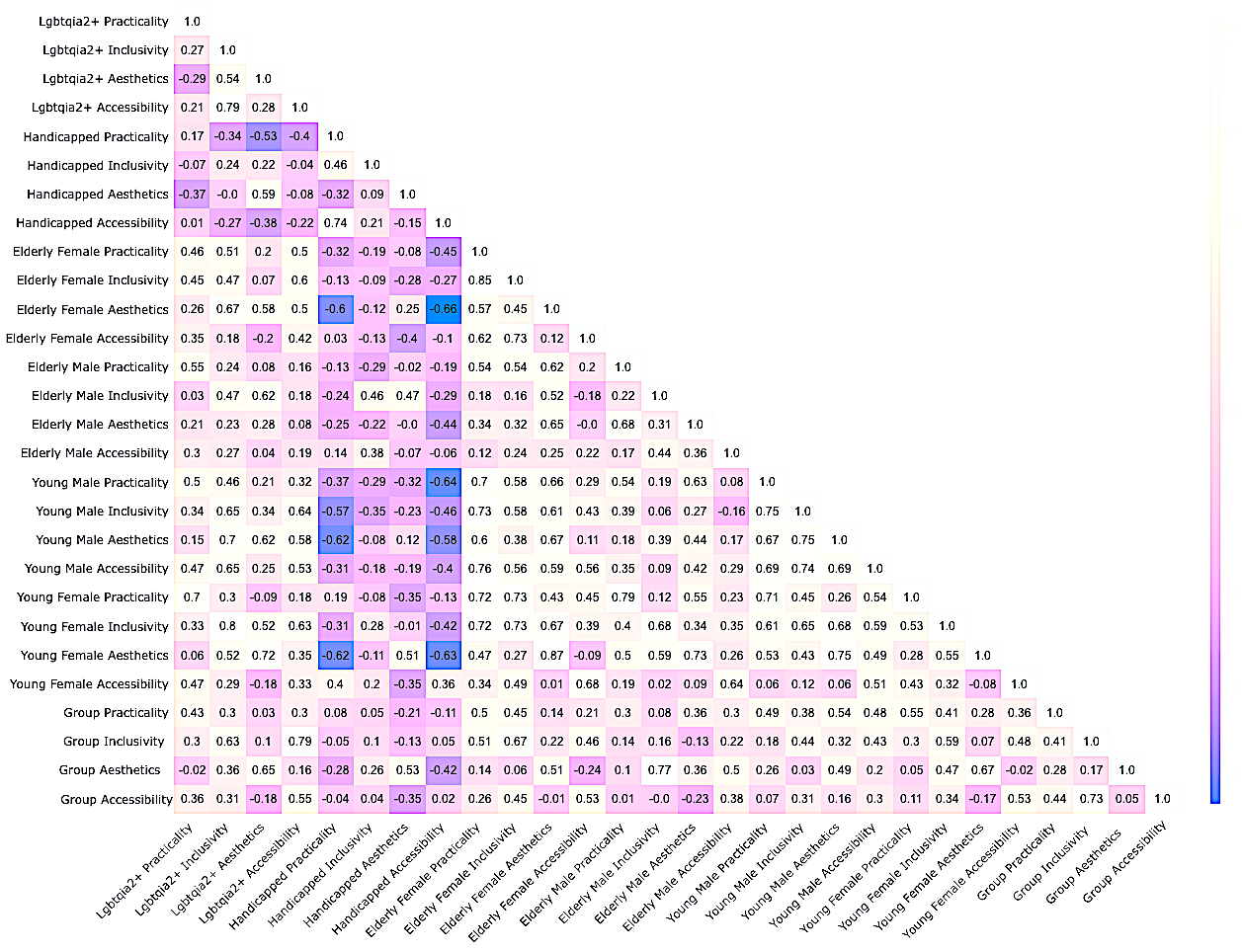}
    \caption{Correlation heatmap for ratings across practicality, inclusivity, aesthetics, and accessibility, segmented by demographic groups. Practicality and accessibility correlate more strongly. Aesthetics and inclusivity vary or invert, reflecting divergent priorities.}
    \Description{Correlation heatmap}
    \label{fig:heatmap_scores_demographic}
\end{figure*}

\subsection{Criteria Rating Table}
\begin{figure*}[htbp]
    \centering
    \includegraphics[width=\textwidth]{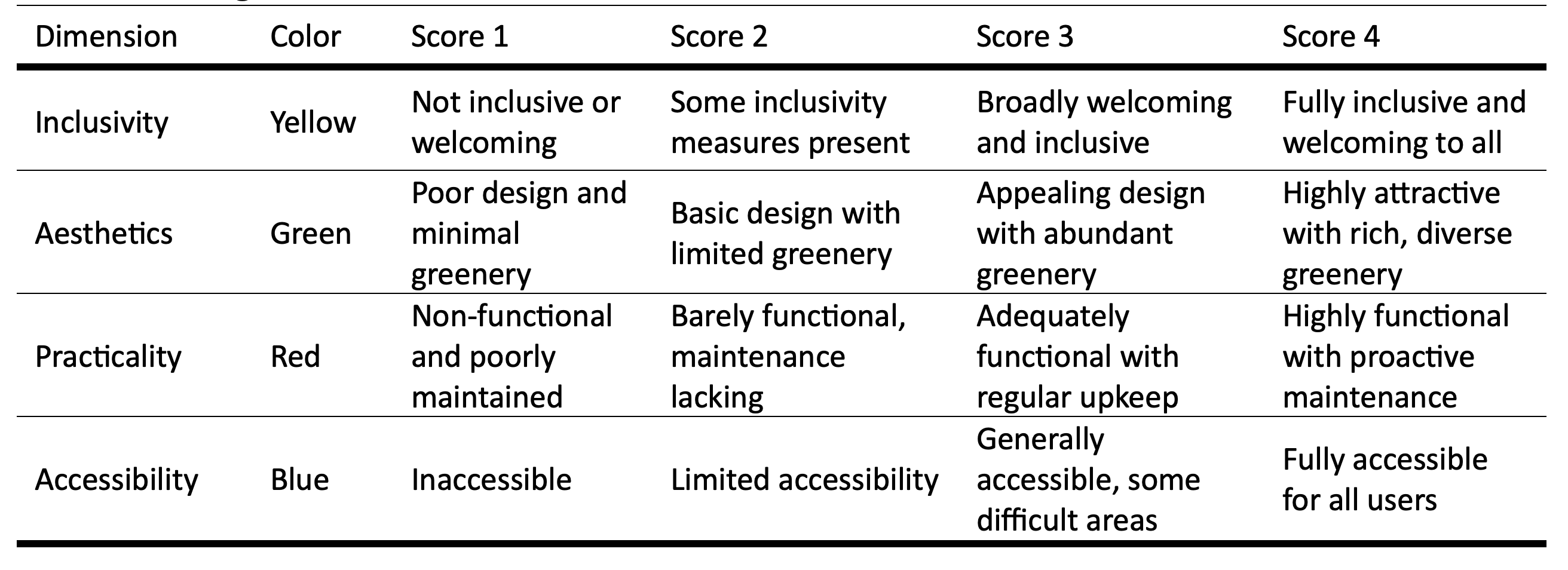}
    \caption{Criteria rating table for four evaluation criteria on a four-level scale. A sticker-based color system supported quick comparison during group sessions.}
    \Description{Criteria Rating Table}
    \label{fig:criteria_rating_table}
\end{figure*}

\subsection{Representative Street Views}
\begin{figure*}[htbp]
    \centering
    \includegraphics[width=\textwidth]{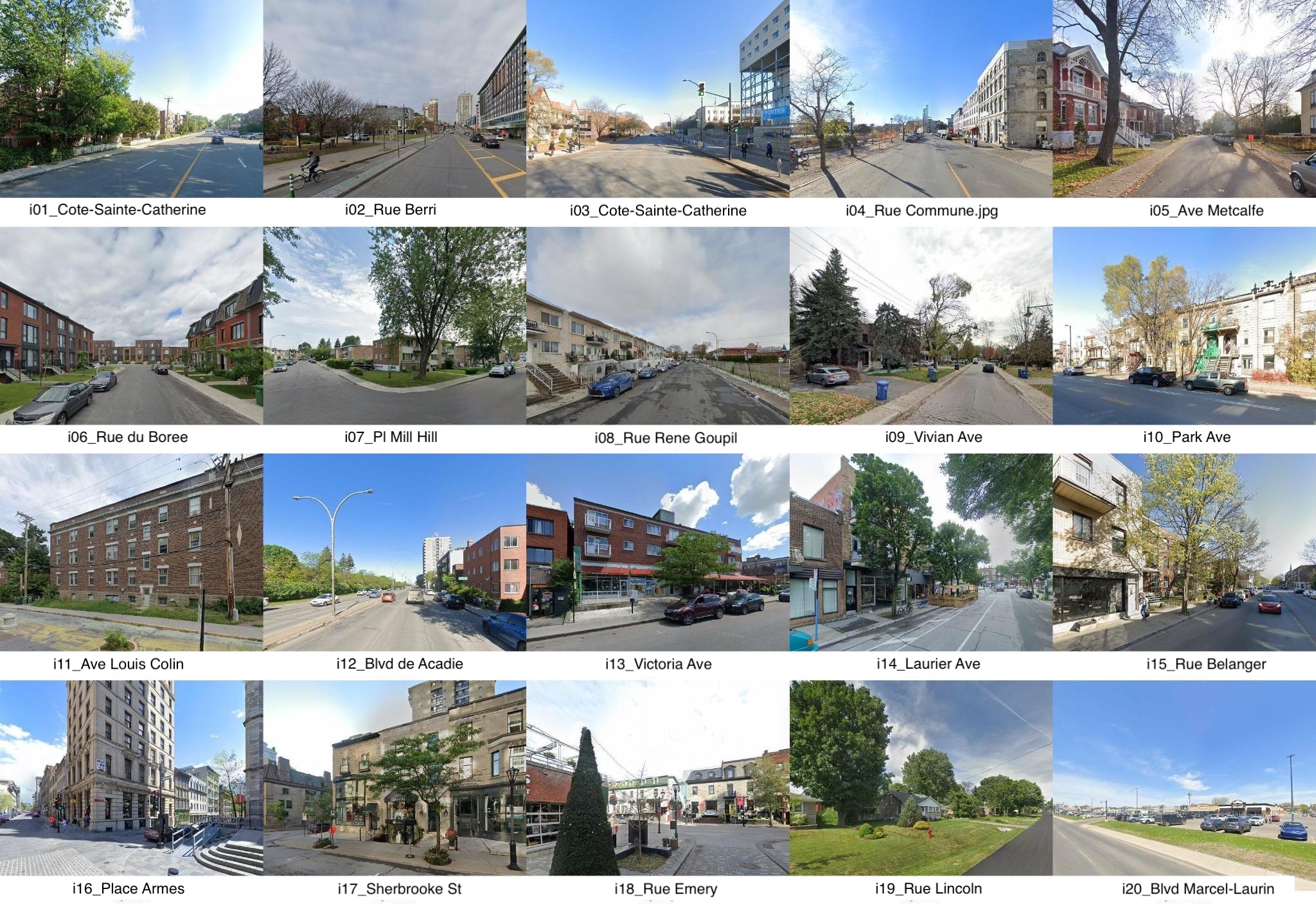}
    \caption{Representative street views across twenty Montreal neighborhoods. The images illustrate variation in land use, pedestrian infrastructure, greenery, and streetscape composition.}
    \Description{Representative Street Views}
    \label{fig:street_views}
\end{figure*}

\subsection{Spatial Distribution of Study Sites}
\begin{figure*}[htbp]
    \centering
    \includegraphics[width=\textwidth]{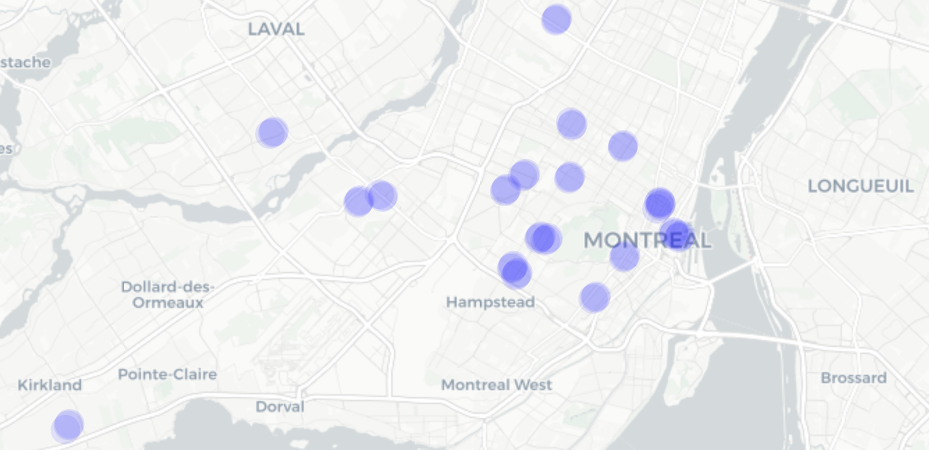}
    \caption{Spatial distribution of twenty sites across Montreal. Locations span socio-economic contexts, land uses, densities, and historical characteristics. Base map: OpenStreetMap.}
    \Description{Spatial Distribution of Study Sites}
    \label{fig:spatial_distribution}
\end{figure*}

\subsection{Street Diversity Classification Guide}
\begin{figure*}[htbp]
    \centering
    \includegraphics[width=\textwidth]{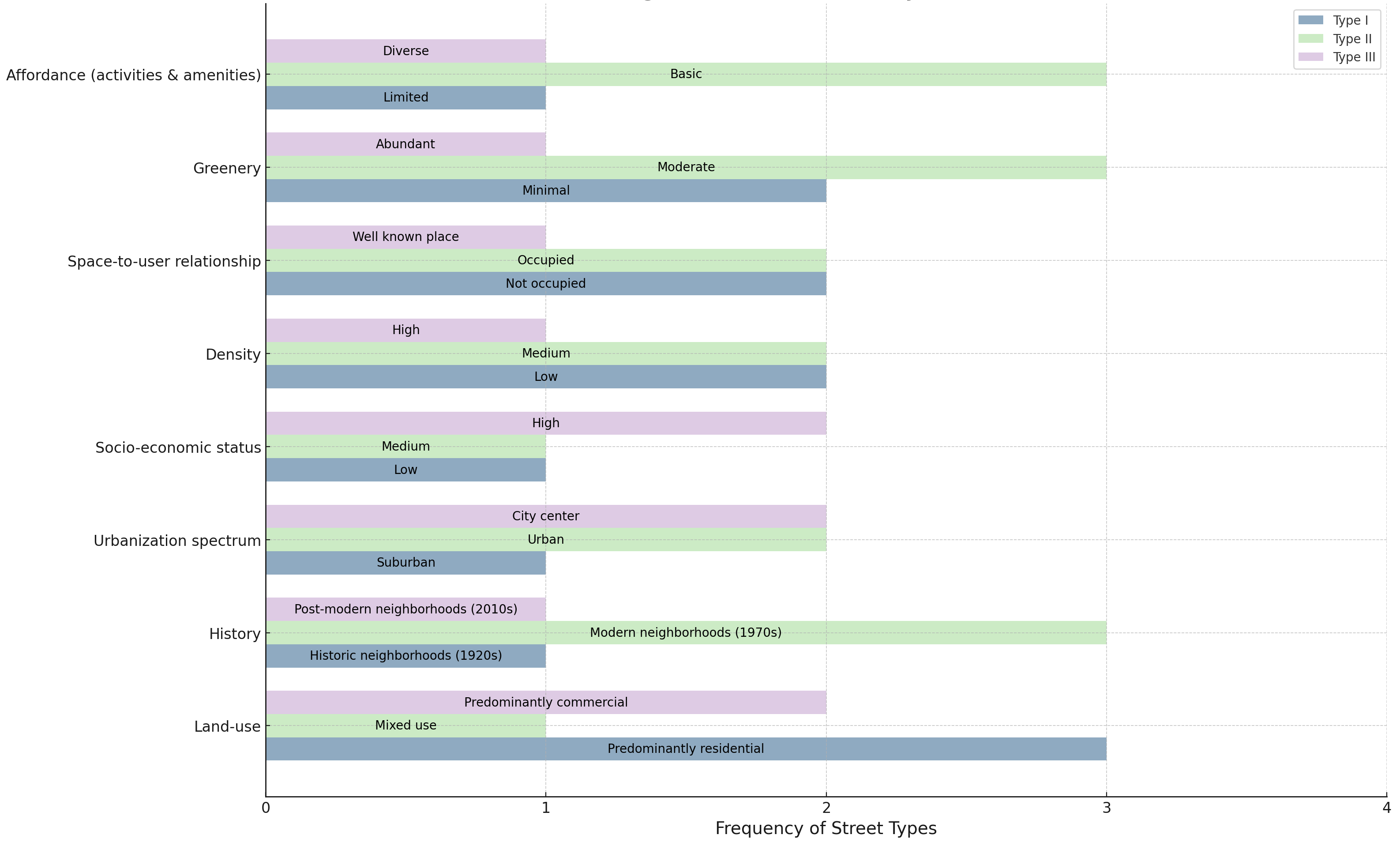}
    \caption{Socio-spatial matrix used to select diverse streets by land use, history, greenery, and aesthetic features.}
    \Description{A matrix categorizing streets}
    \label{fig:street_diversity_classification}
\end{figure*}

\section{Prompts and Hyperparameters}\label{app:prompts}
\paragraph{Decoding.} Temperature 0.4, top-$p$ 0.9, maximum 300 tokens per agent message, maximum 200 tokens for mediator summaries.\\
\paragraph{Global system prompt.} \emph{You are participating in a policy design negotiation. Follow your role card. Keep justifications brief. Do not reveal hidden reasoning. Be civil and discuss only the scenario. Avoid identity-denigrating claims.}\\
\paragraph{Mediator system prompt.} Enforce the budget, maximize the minimum utility, then the sum, provide short public justifications, reject unsafe or off-topic content, and compute utilities only from public cards \cite{mushkani2025righttooverridecriticalurbancontrol}.

\section{Persona Cards (Abridged)}\label{app:personas}
Each card specifies priorities, red lines, and weights $w_{gk}$ that correspond to Section~\ref{subsec:personas}. For example, the wheelchair user prioritizes accessibility, then practicality, then security, followed by inclusivity and aesthetics. A red line requires accessibility of at least four unless the budget can raise it to four within the negotiated package. Weights are Acs 0.50, Pra 0.20, Sec 0.15, Inc 0.10, and Aes 0.05. Other personas are defined analogously and are included in the anonymized repository for artifact review.

\section{Scenario Cards and Improvement Menu}\label{app:scenario-cards}
The pilot used three base designs as in Table~\ref{tab:pilot-roles-scores} and four improvements with costs and score deltas. Ramps with tactile paving increase accessibility by one point at a cost of two. Benches with shade increase practicality by one and inclusivity by 0.2 at a cost of two. Lighting increases security by one at a cost of one. Inclusive signage with art increases inclusivity and aesthetics by one each at a cost of two. Scores cap at five. The budget for the pilot was four.

\section{Pilot Negotiation Transcript}\label{app:transcript}
\begin{quote}\small
\textbf{Mediator}: Budget four. Consider A, B, or C and improvements mA through mD. State one base and up to two improvements.\\
\textbf{W}: Prefer B. Need ramps with tactile paving. Would accept inclusive signage and art.\\
\textbf{Q}: Prefer C. Strongly request inclusive signage and art. Secondary is lighting.\\
\textbf{SF}: Prefer B. Security is strong; prioritize ramps with tactile paving.\\
\textbf{SM}: Prefer B. Agree on ramps with tactile paving. Lighting is optional.\\
\textbf{YF}: Prefer C or B. Support inclusive signage and art. Benches are useful.\\
\textbf{YM}: Prefer B. Support inclusive signage and art.\\
\textbf{Mediator}: Propose B with ramps and inclusive signage and art for total cost four. Attributes become Acs five, Inc four, Aes three, Pra four, Sec four. This maximizes the minimum utility and improves the total. Accept.\\
\textbf{All agents}: Accept.
\end{quote}

\end{document}